\documentclass{jpsj2}

\usepackage{color}

\title{
Theory of Pairing Assisted Spin Polarization in Spin-Triplet Equal Spin Pairing: 
Origin of Extra Magnetization in Sr$_2$RuO$_4$ in Superconducting State
}

\author{
Kazumasa Miyake 
}

\inst
{
Toyota Physical and Chemical Research Institute, Nagakute, Aichi 480-1192, Japan
}
\recdate
{
February 12, 2014
}

\abst
{
It is shown that an extra magnetization is induced by an onset of the equal-spin-pairing 
of spin triplet superconductivity if the energy dependence of the density of states of 
quasiparticles exists in the normal state.  It turns out that the effect is 
observable in  Sr$_2$RuO$_4$ due to the existence of van Hove singularity in the density of 
states near the Fermi level, explaining the extra contribution in the Knight shift reported 
by Ishida {\it et al}\/.  It is also quite non-trivial 
that this effect exists even without external magnetic field, which implies that the time reversal 
symmetry is spontaneously broken in the spin space.   
}

\begin{document}
\sloppy
\maketitle

Properties of the Fermi superfluidity sustained by the triplet pairing have been 
discussed extensively since the discovery of superfluid $^3$He in 1972, and 
its fundamental aspects seem to have been clarified so far.~\cite{Leggett}  
On the other hand, it has recently been measured by the Knight shift that the magnetization 
of Sr$_2$RuO$_4$, which is considered to be a triplet superconductor in the 
equal-spin-pairing (ESP) state,~\cite{Mackenzie} exhibits an extra magnetization under 
the external magnetic 
field other than that expected in the ESP state.~\cite{Ishida}  This phenomenon cannot 
be understood in the framework of spin-singlet pairing state, while some researchers 
doubt the spin-triplet state because the first-order superconducting transition has been 
observed under the magnetic field which is characteristic of the paramagnetic effect in the 
spin-singlet pairing state.~\cite{Yonezawa}  In this sense, it is desired to give a 
theoretical explanation for this phenomenon reported by Ishida.~\cite{Ishida}  
In this Letter we discuss theoretically the mechanism for such an extra magnetization in 
Sr$_2$RuO$_4$, which is considered to be in the spin-triplet ESP 
superconducting state, under the external magnetic field.  This gives an explanation for 
the recent observation of extra magnetization in Sr$_2$RuO$_4$.~\cite{Ishida}   

A physical reason for this extra contribution is rather simple. Under the 
magnetic field, the density of states (DOS) of the normal state quasiparticles of up-spin, 
$N_{\uparrow}(\xi)$, and those of down-spin, $N_{\downarrow}(\xi)$, are different 
if the particle-hole symmetry is apparently broken, i.e., $N(\xi)$'s are not constant 
but have a considerable linear term in the quasiparticle energy $\xi$ measured from the chemical 
potential. Then, the free energy gains associated with Cooper pair condensation 
are different in general, resulting in a redistribution of up-spin and 
down-spin components so as to gain much more condensation energy. Therefore, depending on 
the sign of the linear term of $N(\xi)$, the extra magnetization change arises under the 
external magnetic field $H$.  This mechanism was first predicted almost four decades ago 
by S. Takagi as a 
possible effect of discontinuity in the spin susceptibility of superfluid $^{3}$He at the 
critical temperature $T_{\rm c}$ where $^{3}$He exhibits a second-order phase 
transition from the normal to the A phase at $H=0$.~\cite{Takagi}  The paper by Takagi 
also predicted that in the A1 phase there exists an extra spin-polarization independent 
of $H$ other than the BCS-type contribution. 

On the other hand, Takagi's theory predicted that the extra magnetization quickly fades 
away in the A (or A2) phase where both up- and down-spin components are forming the Cooper pairs. 
This is because Takagi's theory did not take into account the redistribution of fermions 
with up- and down-spin components {\it in the SC state}, while it took into account the migration 
of fermions in the normal down-spin band to the up-spin ESP state in the A1 phase.  Here, we reconsider 
Takagi's discussion and extend it to the ground state under the magnetic field $H$.  

\begin{figure}[h]
\begin{center}
\rotatebox{0}{\includegraphics[width=0.9\linewidth]{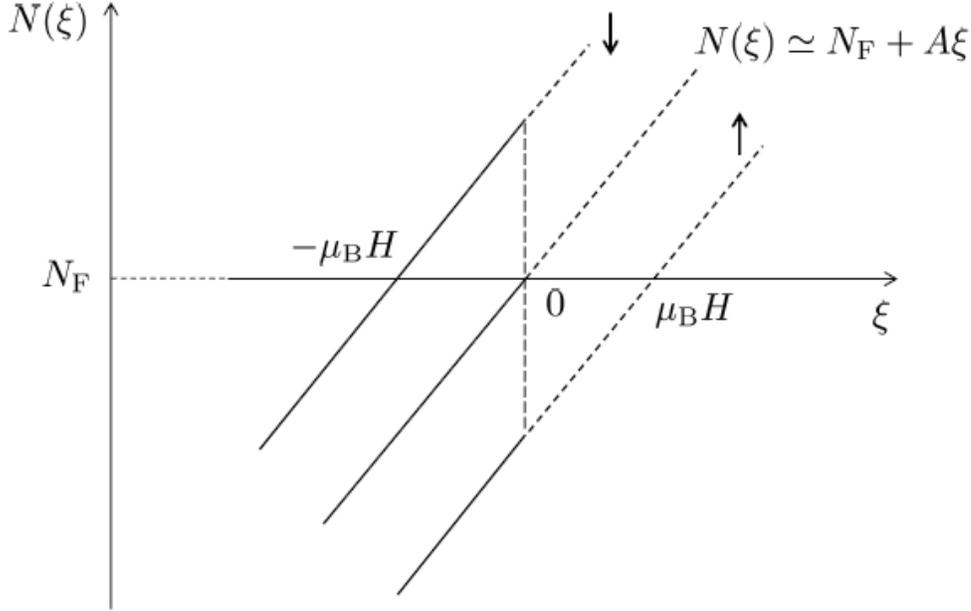}}
\caption{
Density of states $N(\xi)$ vs energy $\xi$ of quasiparticles measure from 
the Fermi level. Line passing $\xi=0$, $\mu_{\rm B}H$, and $-\mu_{\rm B}H$ are 
DOS without magnetic field $H$, for up-spin band, and down-spin 
band, respectively.  Full (dashed) lines indicate the state is occupied (unoccupied). 
Chemical potential shift due to the magnetic field is 
neglected as a negligible effect of the order of 
${\cal O}{[}(\mu_{\rm B}H/\epsilon_{\rm F}^{*})^{2}{]}$. 
}
\label{Fig:1}
\end{center}
\end{figure}

To begin with, we assume that a $\xi$ dependence of the DOS $N(\xi)$ without the magnetic 
field $H$ are given by 
\begin{equation}
N(\xi)\simeq N_{\rm F}+A\xi.
\label{eq:1}
\end{equation}
Then, the DOS of up spin, $N_{\uparrow}(\xi)$, and down-spin, $N_{{\downarrow}}(\xi)$, 
under the field $H$ are shifted as shown in Fig.\ \ref{Fig:1}. Here, we neglect 
the shift in the chemical potential of the order of ${\cal O}(\mu_{\rm B}H/\epsilon_{\rm F}^{*})^{2}$, 
$\epsilon_{\rm F}^{*}$ being the effective Fermi energy of the quasiparticles. 

\subsection*{Ground State}
First, we discuss the case of ground state.  Let us define the difference of condensation energy 
for {majority} down-spin and {minority} up-spin states in the ground state of 
EPS pairing as 
\begin{equation}
\delta E_{\rm cond}={\left[-\frac{1}{2}N_{{\rm F}\downarrow}\Delta_{\downarrow}^{2}-
\left(-\frac{1}{2}N_{{\rm F}\uparrow}\Delta_{\uparrow}^{2}\right)\right]}
\times\frac{1}{2}{,}
\label{eq:2}
\end{equation}
{where $N_{{\rm F}\downarrow}\equiv N_{\rm F}+A\mu_{\rm B}H$ and
$N_{{\rm F}\uparrow}\equiv N_{\rm F}-A\mu_{\rm B}H$, and $\Delta_{\downarrow}$ and 
$\Delta_{\uparrow}$ are the superconducting gap of down-spin and up-spin 
components, respectively. }
With the use of the weak-coupling expression for the superconducting (SC) gap $\Delta$'s, 
$\Delta=\epsilon_{\rm c}^{*}\exp(-1/VN_{\rm F})$, and eq.\ (\ref{eq:1}) for the DOS's, 
$\delta E_{\rm cond}$ is expressed as  
\begin{eqnarray}
& &\delta E_{\rm cond}=-\frac{{1}}{4}
(\epsilon_{\rm c}^{*})^{2}
\left\{(N_{\rm F}+A\mu_{\rm B}H)\exp\left[-\frac{2}{V(N_{\rm F}+A\mu_{\rm B}H)}\right]
\right.
\nonumber
\\
& &\qquad\qquad\qquad\qquad\quad
\left.
-(N_{\rm F}-A\mu_{\rm B}H)\exp\left[-\frac{2}{V(N_{\rm F}-A\mu_{\rm B}H)}\right]
\right\},
\label{eq:3}
\end{eqnarray}
where we have substituted relations $N_{{\rm F}\downarrow}= N_{\rm F} + A\mu_{\rm B}H$ and 
$N_{{\rm F}\uparrow}=N_{\rm F}-A\mu_{\rm B}H$. 
Then, the derivative $\partial\delta E_{\rm cond}/\partial H$ at $H=0$ is given as  
\begin{equation}
\left(\frac{\partial\delta E_{\rm cond}}{\partial H}\right)_{H=0}
=-\frac{N_{\rm F}}{2}\Delta^{2}
\frac{A\mu_{\rm B}}{N_{\rm F}}\left(1+\frac{2}{VN_{\rm F}}\right).
\label{eq:4}
\end{equation}
Therefore, up to the linear term in $H$, the $\delta E_{\rm cond}$ is given as 
\begin{equation}
\delta E_{\rm cond}\simeq 
-\frac{N_{\rm F}}{2}\Delta^{2}
\frac{A\mu_{\rm B}}{N_{\rm F}}\left(1+\frac{2}{VN_{\rm F}}\right)H.
\label{eq:5}
\end{equation}
{
If $A>0$ as shown in Fig.\ \ref{Fig:1}, $\delta E_{\rm cond}<0$, which implies that the 
$\downarrow$-spin pairs have much lower energy than the $\uparrow$-spin ones.  This calculation 
has been performed on the constraint that the distribution of $\downarrow$-spin and 
$\uparrow$-spin electrons number is fixed as the same as {\it in the normal state}.  
However, if this constraint were relaxed, electrons 
forming Cooper pairs should have migrated from $\uparrow$-spin to $\downarrow$-spin band to gain 
more condensation energy, giving rise to an extra magnetization.  

In order to estimate this extra magnetization, we first consider the case without external magnetic field.  
The estimation leading to eq.\ (\ref{eq:5}) is valid also in this case where magnetization 
$\delta m$ increases virtually (associated with migration of Cooper pairs from $\uparrow$-spin 
to $\downarrow$-spin band), if $H$ in eq.\ (\ref{eq:5}) is replaced by 
$\delta m/\chi$, with $\chi$ being the magnetic susceptibility in the normal state.  
Namely, if $A>0$ as shown in Fig.\ \ref{Fig:1}, the virtual magnetization $\delta m$ causes 
energy gain given by eq.\ (\ref{eq:5}) with $H$ replaced by $\delta m/\chi$.  On the other hand, 
the virtual magnetization $\delta m$ is accompanied by energy cost corresponding to the 
magnetic energy $(\delta m)^{2}/2\chi$.  Then, the total energy change $\Delta E(\delta m)$, 
due to this virtual magnetization $\delta m$, is given as 
\begin{equation}
\Delta E(\delta m)\simeq -\frac{N_{\rm F}}{2}\Delta^{2}
\frac{A\mu_{\rm B}}{N_{\rm F}}\left(1+\frac{2}{VN_{\rm F}}\right)\frac{\delta m}{\chi}
+\frac{(\delta m)^{2}}{2\chi}.
\label{eq:6}
\end{equation}
By minimizing this with respect to $\delta m$, we obtain a spontaneous magnetization 
$\delta m$ as 
\begin{equation}
\delta m\simeq \frac{N_{\rm F}}{2}\Delta^{2}
\frac{A\mu_{\rm B}}{N_{\rm F}}\left(1+\frac{2}{VN_{\rm F}}\right).
\label{eq:7}
\end{equation}
Namely, the time reversal symmetry is spontaneously broken even without the magnetic field. 
Of course, negative magnetization $\delta m$ given by 
eq.\ (\ref{eq:7}) with negative sign is also possible, without the external magnetic field, 
as in the case of Ising-like ferromagnetic order.   
In any case, these spontaneously induced magnetizations are caused by the migration of 
Cooper pairs among opposite spin components to gain the condensation energy. 

This induced extra magnetization exists also under the magnetic field $H$. In this case, 
the sign of $\delta m$ is positive, if $A>0$ as in Fig.\ \ref{Fig:1}.  Indeed, 
t}{he total energy $E(m+\delta m)$, where $m$ is the magnetization 
in the conventional ESP state} {under the magnetic field} 
{as discussed below and $\delta m$ is the deviation from the conventional 
one owing to the effect of migration of Cooper pairs, is given as 
\begin{equation}
E(m+\delta m)=E(m)-\frac{N_{\rm F}}{2}\Delta^{2}
\frac{A\mu_{\rm B}}{N_{\rm F}}\left(1+\frac{2}{VN_{\rm F}}\right)\frac{\delta m}{\chi}
+\left[\frac{(m+\delta m)^{2}}{2\chi}-\frac{m^{2}}{2\chi}\right]-\delta m\, H,
\label{eq:7b}
\end{equation}
where} {the first term represents the ``conventional" condensation energy 
under the magnetic field $H$ giving the magnetization $m$,} {the second term the energy 
gain due to the migration of Cooper pairs {causing the change $m\to m+\delta m$}, 
the third term the energy loss due to the excess 
spin polarization, and the last term the excess Zeeman energy under the 
magnetic field $H$.} {The explicit form of the first term $E(m)$ of r.h.s.  
in eq.\ (\ref{eq:7b}) is given by eq.\ (\ref{eq:A1}) as shown below, in which the effect of 
the magnetic field is taken into account only through the difference of the DOS of majority 
and minority bands in the normal state.}  
{The form of the second term of r.h.s. in eq.\ (\ref{eq:7b}) is derived from the 
expression eq.\ (\ref{eq:5}) for the expression of the energy gain by replacing $H$ by a 
"magnetic field" $\delta m/\chi$ corresponding to the excess magnetization $\delta m$.  
By minimizing $E(m+\delta m)$, eq. (\ref{eq:7b}), with respect to $\delta m$, we easily arrive at the 
relation (\ref{eq:7}), considering that the relation $m=\chi H$ holds in the conventional ESP state,} 
{ 
except for a small correction given by eq.\ (\ref{eq:A2}) as shown below.  
The latter correction is of the order of 
${\cal O}{[}(\Delta/\epsilon_{\rm F}^{*})^{2}{]}$  
which gives only a negligibly small correction to $\delta m$, eq.\ (\ref{eq:7}), of the relative 
order of ${\cal O}(\mu_{\rm B}H/\epsilon_{\rm F}^{*})\ll 1$.  
}

The size of coefficient $A$ in eq.\ (\ref{eq:1}) is parameterized as 
$A=N_{\rm F}(a/\epsilon_{\rm F}^{*})$, where $a\sim{\cal O}(1)$ parameterizes 
steepness of the  slope of $N({\xi})$ around $\xi=0$.  The magnetization $m_{\rm n}$ in 
the normal state under the magnetic field $H$ is given by 
$m_{\rm n}\simeq 2\mu_{\rm B}^{2}N_{\rm F}H/(1+F_{0}^{\rm s})$, $F_{0}^{\rm a}$ being the Fermi liquid 
parameter for the correction of the magnetic susceptibility.  
Therefore, the ratio of $\delta m$ and $m_{\rm n}$ is given by 
\begin{equation}
\frac{\delta m}{m_{\rm n}}=\frac{1}{4}
\frac{a\Delta^{2}}{\mu_{\rm B}H \epsilon_{\rm F}^{*}}
\left(1+\frac{2}{VN_{\rm F}}\right)(1+F_{0}^{\rm a}).
\label{eq:8}
\end{equation}

There exists other ``conventional contribution" to the magnetization through the $H$ dependence 
of the condensation energy $E_{\rm cond}$ in the ground state as discussed in ref.\ \citen{Mineev} 
in some different context.  Here, ``conventional contribution" implies that obtained without 
migration of Cooper pairs of down- and up-spin components.  
Indeed, the $E_{\rm cond}$ is given by 
\begin{eqnarray}
& &E_{\rm cond}=-\frac{{1}}{4}
(\epsilon_{\rm c}^{*})^{2}
\left\{(N_{\rm F}+A\mu_{\rm B}H)\exp\left[-\frac{2}{V(N_{\rm F}+A\mu_{\rm B}H)}\right]
\right.
\nonumber
\\
& &\qquad\qquad\qquad\qquad\quad
\left.
+(N_{\rm F}-A\mu_{\rm B}H)\exp\left[-\frac{2}{V(N_{\rm F}-A\mu_{\rm B}H)}\right]
\right\}.
\label{eq:A1}
\end{eqnarray}
Then, the magnetization $m_{\rm s}\equiv -(\partial E_{\rm cond}/\partial H)$ 
(at $H\not=0$) is calculated as 
\begin{equation}
m_{\rm s}=\frac{N_{\rm F}}{2}\Delta^{2}
\frac{A\mu_{\rm B}}{N_{\rm F}}
\frac{4}{(VN_{\rm F})^{2}}
\frac{A\mu_{\rm B}H}{N_{\rm F}}, 
\label{eq:A2}
\end{equation}
where the terms of the order of ${\cal O}{[}(A\mu_{\rm B}H/N_{\rm F})^{2}{]}$ 
have been discarded. 
This $m_{\rm s}$ is smaller than $\delta m$, eq.\ (\ref{eq:7}),  by a small factor 
$A\mu_{\rm B}H/N_{\rm F}=a\,\mu_{\rm B}H/\epsilon_{\rm F}^{*}\ll 1$.  Therefore, 
the ``conventional contribution", eq.\ (\ref{eq:A2}), can be safely neglected.

\subsection*{GL Region}
Next, we discuss the case in GL region,in which we estimate the free energy gain 
$\delta F$ due to SC condensation in stead of the ground state energy at $T=0$~K.   
In the GL region, the free energy difference 
$\delta F_{\rm cond}\equiv F_{\rm cond}^{(+)}-F_{\rm cond}^{(-)}$ is 
given as follows:~\cite{Leggett}
\begin{equation}
\delta F_{\rm cond}=-\frac{K}{4}\left[
(N_{\rm F}+A\mu_{\rm B}H)\left(T_{\rm c}^{(+)}-T\right)^{2}
-(N_{\rm F}-A\mu_{\rm B}H)\left(T_{\rm c}^{(-)}-T\right)^{2}
\right],
\label{eq:9}
\end{equation}
where the SC transition temperatures are given by 
$T_{\rm c}^{(\pm)}={\tilde{\epsilon_{\rm c}}^{*}}\exp[-1/V(N_{\rm F}\pm A\mu_{\rm B}H)]$, and 
$K\equiv 8\pi^{2}/7\zeta(3)\simeq 9.38$, with $\zeta(x)$ being the Riemann $\zeta$ function. 
By calculations similar to the case $T=0$~K, corresponding to eq.\ (\ref{eq:5}), we obtain
\begin{equation}
\delta F_{\rm cond}\simeq 
-\frac{K}{2}\,N_{\rm F}\frac{A\mu_{\rm B}}{N_{\rm F}}
\left[(T_{\rm c}-T)^{2}+\frac{2T_{\rm c}(T_{\rm c}-T)}{VN_{\rm F}}\right]H.
\label{eq:10}
\end{equation}
In the GL region, $T\simeq T_{\rm c}$, the first term in the bracket is neglected compared to 
the second term. 
Then, corresponding to eq.\ (\ref{eq:7}), the extra magnetization $\delta m$ is given as 
\begin{equation}
\delta m\simeq 
K\,N_{\rm F}\frac{A\mu_{\rm B}}{N_{\rm F}}\frac{1}{VN_{\rm F}}
T_{\rm c}(T_{\rm c}-T).
\label{eq:11}
\end{equation}
Therefore, corresponding to eq.\ (\ref{eq:8}), we obtain the ratio of 
$\delta m$ and $m_{\rm n}$ as 
\begin{equation}
\frac{\delta m}{m_{\rm n}}=
\frac{8\pi^{2}}{7\zeta(3)}\frac{a T_{\rm c}(T_{\rm c}-T)}{\mu_{\rm B}H \epsilon_{\rm F}^{*}}
\frac{1}{VN_{\rm F}}(1+F_{0}^{\rm a})
\label{eq:12}
\end{equation}

The result (\ref{eq:11}) is consistent with that for the extra magnetization in the A1 phase, 
eq.\ (4), predicted in Takagi's paper,~\cite{Takagi}
\begin{equation}
M_{\rm I}-M_{\rm n}=N_{\rm F}T_{\rm c}\mu_{\rm B}\eta(t+\eta h)/2\beta,
\label{Takagi}
\end{equation}
considering that correspondence of parameters between {Takagi's} paper and ours is as follows: 
$t=(T_{\rm c}-T)/T_{\rm c}$, $\eta=AT_{\rm c}/V(N_{\rm F})^{2}$, $h=\mu_{\rm B}H/T_{\rm c}$, 
and that our theory has not taken into account the feed back effect; i.e., $(1/\beta)=K$.  
A difference in overall factor by 2 can be understood from the fact that {Takagi's} eq.\ (4) is 
for near the A1 transition associated with only up-spin pairing while our result eq.\ (\ref{eq:11}) 
is for both up- and down-spin pairings.  The reason why the extra 
magnetization which is independent of the external magnetic field $H$ ($h$) 
is missing in the A2 phase in Takagi's expression, eq.\ (5), seems to be traced back to the fact 
that he has not taken into account the migration of electrons from down-spin to up-spin 
Cooper pairs
in the A phase while he has taken into account that from the down-spin {$^{3}$He nuclei} 
in the normal state to the up-spin Cooper pairs in the A1 phase.  

The ``conventional contribution" to the magnetization through the $H$ dependence 
of the free energy $F_{\rm cond}$ in GL region is calculated similarly to the 
case in the ground state.  The $F_{\rm cond}$ is given as
\begin{equation}
F_{\rm cond}=-\frac{K}{4}\left[
(N_{\rm F}+A\mu_{\rm B}H)\left(T_{\rm c}^{(+)}-T\right)^{2}
+(N_{\rm F}-A\mu_{\rm B}H)\left(T_{\rm c}^{(-)}-T\right)^{2}
\right],
\label{eq:10a}
\end{equation}
Then, the magnetization $m_{\rm s}\equiv -(\partial F_{\rm cond}/\partial H)$ 
(at $H\not=0$) is calculated as 
\begin{equation}
m_{\rm s}\simeq 
K\,N_{\rm F}\frac{A\mu_{\rm B}}{N_{\rm F}}\frac{1}{(VN_{\rm F})^{2}}
\left[2T_{\rm c}(T_{\rm c}-T)+TT_{\rm c}\right]\frac{A\mu_{\rm B}H}{{N_{\rm F}}}, 
\label{eq:11a}
\end{equation}
where the terms of the order of ${\cal O}(A\mu_{\rm B}H/N_{\rm F})^{2}$ have been discarded 
as in the case of ground state above.  This $m_{\rm s}$ is smaller than $\delta m$, 
eq.\ (\ref{eq:11}), by a small factor 
$A\mu_{\rm B}H/N_{\rm F}=a\,\mu_{\rm B}H/\epsilon_{\rm F}^{*}\ll 1$.  Therefore, 
the ``conventional contribution", eq.\ (\ref{eq:11a}), can be safely neglected again.  

It is remarked that the expression (\ref{eq:11a}) is exactly the same as eq.\ (5) 
in Takagi's paper for the A2 phase to the zeroth order in ($T-T_{\rm c}$):~\cite{Takagi}  
\begin{equation}
M_{\rm II}-M_{\rm n}=N_{\rm F}T_{\rm c}\mu_{\rm B}\eta^{2}h/\beta,
\label{Takagi2}
\end{equation}
considering again that correspondence of parameters between {Takagi's} paper and ours is as follows: 
$t=(T_{\rm c}-T)/T_{\rm c}$, $\eta=AT_{\rm c}/V(N_{\rm F})^{2}$, $h=\mu_{\rm B}H/T_{\rm c}$, 
and that our theory has not taken into account the so-called feed back effect 
due to spin fluctuations; i.e., $(1/\beta)=K$ and $\delta=0$.

\subsection*{Order Estimation}
Here we give a rough order estimation for $\delta m/m_{\rm n}$ in Sr$_2$RuO$_4$.  
With the use of the correlation length at $T=0$~K, $\xi_{0}\simeq 1050$~\AA,~\cite{Miyake} 
the effective Fermi energy of the quasiparticles $\epsilon_{\rm F}^{*}$ is estimated as 
\begin{equation}
\epsilon_{\rm F}^{*}\simeq 2.5\times 10^{3}T_{\rm c}\simeq 3.8\times 10^{3}\,{\rm K}.
\label{eq:12a}
\end{equation}
Assuming ${\tilde \epsilon_{\rm c}}\sim \epsilon_{\rm F}^{*}$, the couping constant $VN_{\rm F}$ 
is estimated as 
\begin{equation}
\frac{1}{VN_{\rm F}}\simeq 7.
\label{eq:13}
\end{equation}
The SC gap at $T=0$~K is estimated by using the BCS relation:
\begin{equation}
\Delta\simeq 1.7\times T_{\rm c}\simeq 2.6\,{\rm K}.
\label{eq:14}
\end{equation}
The Landau parameter $F_{0}^{\rm a}$ is estimated from the Wilson ratio as 
$F_{0}^{\rm a}\simeq -0.5$.~\cite{Maeno}

The magnetic field $H\simeq 1$~{T}, used in the NMR Knight shift measurements, 
is equivalent to $H\mu_{\rm B}\simeq 0.67$~K. 
Then, the ratio $\delta m/m_{\rm n}$, eq.\ (\ref{eq:8}), at $T=0$~K is estimated as 
\begin{equation}
\frac{\delta m}{m_{\rm n}}\simeq 5.0\times 10^{-3}\times a.
\label{eq:15}
\end{equation}
Since there exists the van Hove singularity in the DOS of the $\gamma$ band 
just above the Fermi level, the parameter $a$, parameterizing the steepness 
of the slope in DOS at the Fermi level, can be much larger than 1/2, the value for free fermions. 
Indeed, according to Fig.\ 41 for the DOS of $\gamma$ band in ref.\ \citen{Bergemann}, and 
considering $m^{*}/m_{\rm band}\simeq 5.5$,~\cite{Mackenzie2} 
the parameter $a$ is estimated as $a\simeq 3.6$. 
The effect of the $\alpha$ and $\beta$ bands may give some additional contribution. 
However, since the DOS of the $\gamma$ band dominates those of $\alpha$ and 
$\beta$ bands, the effect is expected to be limited.  Thus, the ratio $\delta m/m_{\rm n}$ at 
$T=0$~K can be a few \% in consistent with the Knight shift measurements reported 
in ref.\ \citen{Ishida}, while the above estimations are rather crude.

\subsection*{Discussions}
It is noted that the excess magnetization given by eqs.\ (\ref{eq:7}) and (\ref{eq:11}) 
exists without external magnetic field.  This implies that such a magnetization gives a 
spontaneous magnetic field breaking time reversal symmetry.  It is crucial that this effect is 
not related to the orbital effect of degenerate component of the Cooper pairs, such as 
$(\sin k_{x}+{\rm i}\sin k_{y})$ state.~\cite{Miyake2}  
The size of this magnetic field is roughly estimated as follows: By using the relation 
$N_{\rm F}=3N/4\epsilon_{\rm F}^{*}$ for a free dispersion, eq.\ (\ref{eq:7}) is reduced to 
\begin{equation}
\delta m=\frac{3}{8}\left(\frac{\Delta}{\epsilon_{\rm F}^{*}}\right)^{2}a
\left(1+\frac{2}{VN_{\rm F}}\right)N\mu_{\rm B}.
\label{eq:16}
\end{equation}
By assuming that there exists one electron per unit cell 
($a=b=3.9\times 10^{-10}$\,m, and $c=(12.7/2)\times 10^{-10}$\, m), 
the number of electrons $N$ per unit volume is estimated as $N\simeq 1.04\times 10^{28}$. 
Then, using the values, eqs.\ (\ref{eq:12a}), (\ref{eq:13}), and (\ref{eq:14}), 
$\delta m$ at $T=0$ is estimated as 
\begin{equation}
\delta m\simeq 0.92\,{\rm J}\cdot{\rm T}^{-1}. 
\label{eq:17}
\end{equation}
This corresponds to the magnetic field $\delta B$ as
\begin{equation}
\delta B=\mu_{0}\delta m\simeq 1.1\times 10^{-6}\,{\rm T}=1.1\times 10^{-2}\,{\rm G}, 
\label{eq:18}
\end{equation}
where $\mu_{0}=4\pi\times 10^{-7}\,$H$\cdot$m$^{-1}$ is the magnetic permeability of vacuum.  
This is far smaller than the lower critical field $H^{ab}_{{\rm c}1}=10\,{\rm G}$ and 
$H^{c}_{{\rm c}1}=50\,{\rm G}$,~\cite{Akima} so that it would be fully screened out  
by the Meissner effect.  Therefore, it seems technically impossible to observe this small 
spontaneous magnetic field {\it if} the domain size is larger than the penetration depth of 
magnetic field.

It is interesting that the effect similar to that observed in Sr$_2$RuO$_4$ 
seems to have been observed in UPt$_3$ although the effect is smaller than that in 
Sr$_2$RuO$_4$ by one order of magnitude.~\cite{Tou}  
It is also interesting that an upper bound of spontaneous magnetic filed of the order 
of 1$\,$mG, one order smaller than a value given by eq.\ (\ref{eq:17}), was reported in UPt$_{3}$ 
on a measurement by using a SQUID magnetometer.~\cite{Kambara}  This is consistent with 
the fact that $\mu$SR measurement of high quality single crystal has given estimations of 
upper bound of the spontaneous magnetization as $\sim 30\,$mG~\cite{Reotier} or 
$\sim 80\,$mG~\cite{Higemoto}.  

The pairing assisted spin polarization should exist also in the A-phase of superfluid $^3$He.  
Indeed, $\delta m/m_{\rm n}$, eq.\ (\ref{eq:8}), is estimated under a hypothetical situation, i.e.,
$T=0$~K and $H=1$~Tesla, as follows: With the use of a parameter set for $^3$He at p = 27 bar 
($\epsilon_{\rm F}^{*}\simeq 1.09$~K, $\Delta=1.7\,T_{\rm c}\simeq 4.3$~mK, 
$\mu_{\rm N}\simeq 1.1\times 10^{-26}$~J/{T}, $1/VN_{\rm F}\simeq 6$, $F_{0}^{\rm a}\simeq -0.755$ 
and $a=1/2$)~\cite{Wheatley}, the ratio $\delta m/m_{\rm n}$ is estimated as  
\begin{equation}
\frac{\delta m}{m_{\rm n}}\simeq 7.7\times 10^{-3}.
\label{eq:19}
\end{equation}
Thus, the extra magnetization in the A-phase of superfluid $^3$He is nearly the same order as that 
expected in Sr$_2$RuO$_4$.

\subsection*{Conclusion}
It has been shown that the extra magnetization (or spin polarization) is induced 
in the ESP state due to the migration of the Cooper pairs from minority to majority 
pairing state to gain the condensation energy (free energy).  This effect seems 
to have been overlooked for four decades, and to give a semi-quantitative 
explanation for the effect which was discovered quite recently by the Knight 
shift measurements in Sr$_2$RuO$_4$ by Ishida and coworkers.  This extra magnetization is 
induced spontaneously even without the external magnetic field.   

\subsection*{Acknowledgments}
The author is grateful to K. Ishida who directed his attention to the present 
problem and stimulating discussions, to Y. Maeno for enlightening discussions, 
and to W. Higemoto, N. Nishida, and A. Sumiyama for informative communications.  
This work is supported by a Grant-in-Aid for Scientific 
Research on Innovative Areas ``Topological Quantum Phenomena" 
(No.22103003) from the Ministry of Education, Culture, Sports, Science and
Technology of Japan, and by a Grant-in-Aid for Scientific 
Research (No.25400369) from the Japan Society for the Promotion of Science.

\end{document}